\documentclass[prl,twocolumn,showpacs,preprintnumbers,amsmath,amssymb,superscriptaddress]{revtex4}

\usepackage{graphicx}
\usepackage{pifont} 
\usepackage{wasysym} 

\begin{document}

\title{Enhancement of quasiparticle recombination in Ta and Al superconductors\\ by implantation of magnetic and nonmagnetic atoms}

\author{R. Barends}
\affiliation{Kavli Institute of NanoScience, Faculty of Applied
Sciences, Delft University of Technology, Lorentzweg 1, 2628 CJ
Delft, The Netherlands}

\author{S. van Vliet}
\affiliation{Kavli Institute of NanoScience, Faculty of Applied
Sciences, Delft University of Technology, Lorentzweg 1, 2628 CJ
Delft, The Netherlands}

\author{J. J. A. Baselmans}
\affiliation{SRON Netherlands Institute for Space Research,
Sorbonnelaan 2, 3584 CA Utrecht, The Netherlands}

\author{S. J. C. Yates}
\affiliation{SRON Netherlands Institute for Space Research,
Sorbonnelaan 2, 3584 CA Utrecht, The Netherlands}

\author{J. R. Gao}
\affiliation{Kavli Institute of NanoScience, Faculty of Applied
Sciences, Delft University of Technology, Lorentzweg 1, 2628 CJ
Delft, The Netherlands}

\affiliation{SRON Netherlands Institute for Space Research,
Sorbonnelaan 2, 3584 CA Utrecht, The Netherlands}

\author{T. M. Klapwijk}
\affiliation{Kavli Institute of NanoScience, Faculty of Applied
Sciences, Delft University of Technology, Lorentzweg 1, 2628 CJ
Delft, The Netherlands}

\date{\today}

\begin{abstract}
The quasiparticle recombination time in superconducting films,
consisting of the standard electron-phonon interaction and a yet to
be identified low temperature process, is studied for different
densities of magnetic and nonmagnetic atoms. For both Ta and Al,
implanted with Mn, Ta and Al, we observe an increase of the
recombination rate. We conclude that the enhancement of
recombination is not due to the magnetic moment, but arises from an
enhancement of disorder.
\end{abstract}

\pacs{74.25.Nf, 74.40.+k, 74.78.Db}

\maketitle

\begin{figure}[b!]
    \centering
    \includegraphics[width=0.85\linewidth]{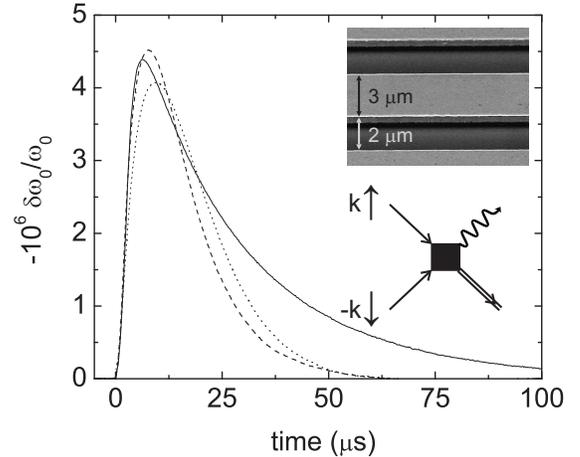}
    \caption{The evolution of the resonance frequency in response to an optical pulse (2 $\mu$s duration)
    of a Ta sample (solid line), Ta implanted with 100 ppm Mn (dashed) and 100 ppm Ta (dotted) (average of 100 traces).
    The initial rise is due to the response time of the resonator,
    the subsequent exponential decay (Ta: $\tau$=28 $\mu$s , Ta with Mn: $\tau$=11 $\mu$s, Ta with Ta: $\tau$=11 $\mu$s)
    reflects the recovery of the equilibrium state (Eq. \ref{equation:f0}).
    The relaxation is due to recombination of quasiparticles into Cooper pairs (depicted in the lower inset).
    A scanning electron micrograph of the coplanar waveguide geometry of the resonator is shown in the upper inset,
    the width of the central line is 3 $\mu$m and the width of the slits is 2 $\mu$m.}
    \label{figure:fig1}
\end{figure}

When a superconductor is perturbed, the equilibrium state is
recovered by the recombination of excess quasiparticle excitations.
Recombination is a binary reaction, quasiparticles with opposite
wave vector and spin combine and join the superconducting condensate
formed by the Cooper pairs, pairs of time-reversed electron states;
the energy is transferred to the lattice by the material-dependent
electron-phonon interaction \cite{kaplan} (symbolically represented
by the lower inset of Fig. \ref{figure:fig1}). With decreasing bath
temperature the number of thermal quasiparticle excitations
available for recombination reduces, and consequently the
recombination time increases exponentially. There is however a
discrepancy between this theory and experiments performed at low
temperatures \cite{barends}. We have found that the relaxation
saturates at low temperatures in both Ta and Al, indicating the
presence of a second physical process which dominates low
temperature relaxation. The energy flux in hot electron experiments
suggests the same pattern \cite{timofeev}.

In the normal state it has become clear that a dilute concentration
of magnetic atoms significantly enhances the inelastic scattering
among quasiparticles \cite{anthore,huard}. In a superconductor the
magnetic moment of the impurity leads to time-reversal symmetry
breaking by spin-flip scattering, altering the superconducting
state. The critical temperature $T_c$ and energy gap $\Delta$
decrease with increasing impurity concentration \cite{abrikosov}.
Depending on the magnetic atom and the host, localized impurity
bound states as well as a band of states within the energy gap can
appear \cite{yazdani,bauriedl,dumoulin}. In order to test the
influence of magnetic impurities on the inelastic interaction in
superconducting films we have implanted both magnetic and
nonmagnetic atoms and measured the relaxation times at temperatures
far below the critical temperature.

We use the complex conductivity $\sigma_1-i \sigma_2$ to probe the
superconducting state. The real part, $\sigma_1$, reflects the
conduction by quasiparticles while the imaginary part, $\sigma_2$,
arises from the accelerative response of the Cooper pairs,
controlling the high frequency ($\omega$) response of the
superconductor \cite{mattis}. The restoration of the equilibrium
state is measured by sensing the complex conductivity while applying
an optical photon pulse. To this end, the superconducting film is
patterned into planar quarter and half wavelength resonators,
comprised of a meandering coplanar waveguide (CPW) with a central
line, 3 $\mu$m wide, and metal slits, 2 $\mu$m wide, see upper inset
Fig. \ref{figure:fig1}, for details see Refs. \cite{barends,day}.
The condensate gives rise to a kinetic inductance $L_k \sim 1/d
\omega \sigma_2$, with $d$ the thin film thickness, which controls
the resonance frequency: $\omega_0 = 2 \pi /4l \sqrt{(L_g+L_k)C}$
for a quarterwave resonator with length $l$, $L_g$ the geometric
inductance and $C$ the capacitance per unit length. Lengths of
several millimeters are used, corresponding to resonance frequencies
of typically 3-6 GHz. The resonators are capacitively coupled to a
feedline. Upon optical excitation the complex conductivity reflects
the change in the quasiparticle density $n_{qp}$ by: $\delta
\sigma_2/\sigma_2 = -\frac{1}{2}\delta n_{qp} / n_{cp}$, with
$n_{cp}$ the Cooper pair density ($n_{qp} \ll n_{cp}$). The
resonance frequency directly senses the variation in the
superconducting state,
\begin{equation}
\label{equation:f0} \frac{\delta \omega_0}{\omega_0} =
\frac{\alpha}{2} \frac{\delta \sigma_2}{\sigma_2} \Big( f(E),\Delta
\Big),
\end{equation}
with $f(E)$ the distribution of quasiparticles over the energy and
$\alpha$ the fraction of the kinetic to total inductance.

The resonators are made from Ta and Al. The Ta film, 280 nm thick,
is sputter-deposited onto a hydrogen passivated, high resistivity
($> 10$ k$\Omega$cm) (100)-oriented Si substrate. A 6 nm Nb seed
layer is used underneath the Ta layer to promote growth of the
desired body-centered-cubic phase \cite{face}. The film critical
temperature is 4.4 K, the low temperature resistivity ($\rho$) is
8.8 $\mu\Omega$cm and the residual resistance ratio ($RRR$) is 3.2.
The Al film, with a thickness of 100 nm, is sputtered onto a similar
Si substrate ($T_c$=1.2 K, $\rho$=0.81 $\mu\Omega$cm and $RRR$=4.5).
Patterning is done using optical lithography, followed by reactive
ion etching for Ta and wet etching for Al. After patterning various
concentrations of Mn, as magnetic atom, and Ta and Al have been
ion-implanted. The Ta film has been implanted with Mn, Ta and Al at
energies of 500, 500 and 250 keV respectively, and the Al film has
been implanted with Mn and Al at 60 and 30 keV, to place the peak of
the concentration near the middle of the film \cite{SRIM}. The Ta
samples are placed on a He-3 sorption cooler in a He-4 cryostat,
with the sample space surrounded by a superconducting magnetic
shield. The Al samples are placed on an adiabatic demagnetization
refrigerator; here a superconducting and cryoperm shield are used.
The optical pulse is provided by a GaAsP (1.9 eV) LED, which is
fibre-optically coupled to the sample box. The transmission of the
feedline near the resonance frequency is sensed using a signal
generator, low noise amplifier and quadrature mixer, allowing for
monitoring the resonance frequency in the time domain
\cite{barends,day}.

\begin{figure}[t!]
    \centering
    \includegraphics[width=1\linewidth]{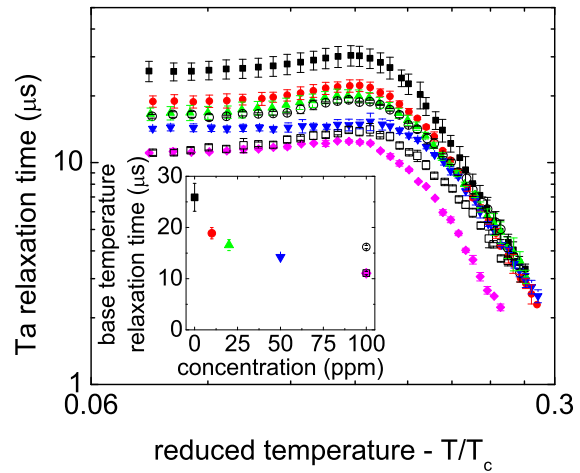}
    \caption{(Color online) The relaxation time as a function of reduced bath temperature in
    Ta ($T_c$=4.4 K)
    with ion-implanted concentrations of Mn: 0 ($\blacksquare$), 10 (\ding{108}), 20 ($\blacktriangle$),
    50 ($\blacktriangledown$) and 100 ppm (\ding{117}), as well as with 100 ppm Ta ($\square$) and 100 ppm Al ($\raise.08\baselineskip\hbox{\fullmoon}$).
    The relaxation times at base temperature (325 mK) are plotted in the inset versus ion concentration.}
    \label{figure:Tadata}
\end{figure}

Typical optical pulse responses are shown in Fig. \ref{figure:fig1}
for Ta quarterwave resonators at the base temperature of 325 mK. The
exponential decrease reflects the restoration of equilibrium in the
superconducting state. The initial rise is due to the response time
of the resonator. The faster decay indicates a faster relaxation for
implanted Ta samples. The temperature dependence of the relaxation
times is shown in Fig. \ref{figure:Tadata} for Ta samples implanted
with a range of concentrations from 0 to 100 ppm Mn, and with 100
ppm Ta and Al. At low temperatures a clear trend of a
\emph{decreasing} relaxation time with \emph{increasing} impurity
concentration is visible, both for samples implanted with Mn as well
as with Ta and Al. Below $T/T_c \sim 0.1$ the relaxation times
become independent of temperature, reaching plateau values of 26
$\mu$s for the unimplanted samples, values down to 11 $\mu$s for
samples implanted with Mn, 11 $\mu$s with Ta and 16 $\mu$s with Al,
clearly decreasing with increasing impurity concentration (see
inset). Near $T/T_c \sim 0.15$ the relaxation times reach a peak
value in all samples. At high temperatures ($T/T_c \gtrsim 0.2 $) we
find that the relaxation times increase with decreasing temperature.
Here, the relaxation times of the implanted samples, except for the
sample with 100 ppm Mn, join with the values of the unimplanted
sample, and is understood as due to the conventional electron-phonon
process \cite{barends}. The critical temperature remains unchanged.

\begin{figure}[t!]
    \centering
    \includegraphics[width=1\linewidth]{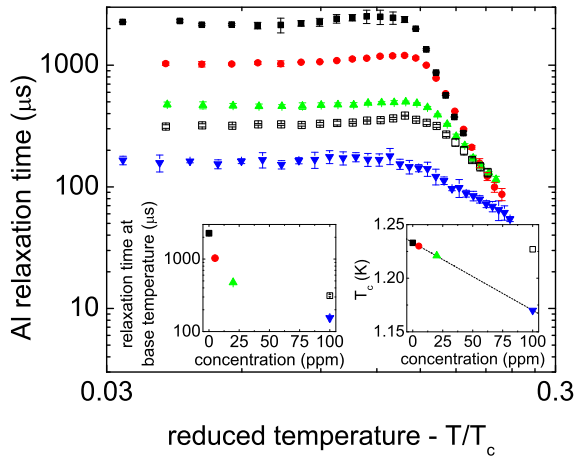}
    \caption{(Color online) The relaxation time as a function of reduced bath temperature in Al
    with various ion-implanted concentrations of Mn: 0 ($\blacksquare$), 5 (\ding{108}), 20
    ($\blacktriangle$) and 100 ppm ($\blacktriangledown$), as well as with 100 ppm Al ($\square$).
    The left inset shows the relaxation time at base temperature versus ion concentration.
    The critical temperature decreases only with increasing Mn concentration (right inset).}
    \label{figure:Aldata}
\end{figure}

In Al samples, halfwave resonators, implanted with 0 to 100 ppm Mn
or 100 ppm Al the relaxation times follow a similar pattern, see
Fig. \ref{figure:Aldata}. The effect of the implanted impurities is
most significant at the lowest temperatures (below $T/T_c \sim
0.1$), where the plateau value of the relaxation time is decreased
by an order of magnitude: from a value of 2.3 ms for unimplanted Al
down to 320 $\mu$s for Al with 100 ppm Al and 150 $\mu$s for Al with
100 ppm Mn (see left inset). A slight nonmonotonic temperature
dependence is observed for all samples. Above $T/T_c \gtrsim 0.2$
the relaxation times increase with decreasing temperature. In
addition, the sample critical temperature decreases linearly with
increasing Mn concentration, see right inset of Fig.
\ref{figure:Aldata}, with $\Delta T_c /\Delta c_{\mathrm{Mn}} =
-0.63$ mK/ppm (dashed line), while remaining unchanged when
implanting Al.

We interpret the relaxation as due to the recombination of
quasiparticles near the gap energy: First, we probe $\sigma_2$ which
is associated with the Cooper pairs. Second, identical relaxation
times are found when creating quasiparticle excitations near the gap
energy by applying a microwave pulse at the resonance frequency
$\omega_0$. In addition, the data are not influenced by
quasiparticle outdiffusion as no length dependence was observed in
the Al half wavelength and Ta quarter wavelength resonators used.
Moreover, the relaxation time is independent of the photon flux for
the small intensities used. Furthermore, the samples are well
isolated from thermal radiation: we observe no significant change in
relaxation time when varying the temperature of the cryostat or of a
blackbody placed next to the sample box. Finally, the significant
effect of the implantation of impurities indicates that the
relaxation time reflects the restoration of equilibrium in the
superconducting films.

The data show a clear trend of decreasing relaxation time in both Ta
and Al with an increasing ion-implanted impurity concentration. The
significant decrease at the lowest temperatures indicates that the
dominant low temperature relaxation channel is enhanced while the
relaxation process at higher temperatures is less affected.

In a superconductor the magnetic nature of the atom depends on the
coupling between its spin and the host conduction electrons. Mn has
been shown to retain its magnetic moment in Nb, V \cite{roy} and Pb
\cite{bauriedl}, acting as pair breaker and giving rise to subgap
states. On the other hand, when Mn is placed inside Al $s-d$ mixing
occurs: the localized $d$ electron states of the transition metal
impurity strongly mix with the conduction band, resulting in the
impurity effectively loosing its magnetic moment as well as an
increase in the Coulomb repulsion \cite{kaiser}. It acts
predominantly as pair weakener: suppressing superconductivity, yet
contrary to the case of pair breaking, showing no evidence of subgap
states \cite{oneil}.

\begin{figure}[b!]
    \centering
    \includegraphics[width=1\linewidth]{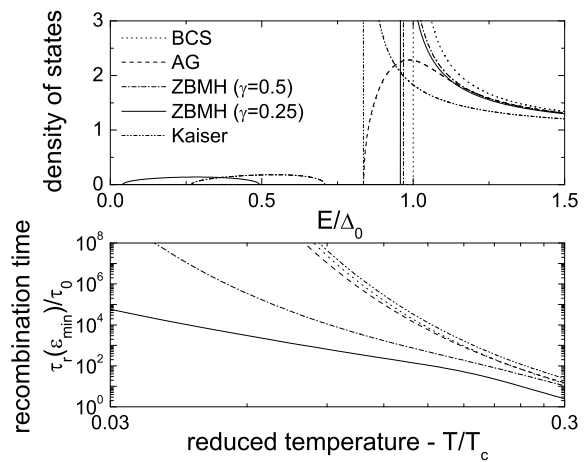}
    \caption{Upper figure: Normalized quasiparticle density of states in the presence of magnetic impurities
    according to pair breaking theories by Abrikosov and Gorkov (AG) as well as Zittartz, Bringer and M\"{u}ller-Hartmann (ZBMH) ($\Gamma/\Delta_0$=0.03)
    and the pair weakening theory by Kaiser (for $\Delta$ identical to the AG case).
    Lower figure: the corresponding recombination times, using Eq. \ref{equation:kaplan}.}
    \label{figure:MHZB}
\end{figure}

In order to quantify a possible influence of magnetic impurities on
recombination we use the conventional theories by Zittartz, Bringer
and M\"{u}ller-Hartmann \cite{zittartz} and Kaiser \cite{kaiser}. In
the presence of pair-breaking impurity bound states develop within
the energy gap near reduced energy $\gamma$. The quasiparticle
excitations, denoted by the Green's function $G$, and the paired
electrons, $F$, are described by: $E=u ( \Delta + \Gamma
\frac{\sqrt{1-u^2}}{u^2-\gamma^2} )$, with
$G(E)=u(E)/\sqrt{u(E)^2-1}$, $F(E)=i/\sqrt{u(E)^2-1}$ and
$\Gamma=\hbar/\tau_{sf}$ the pair-breaking parameter. For $\gamma
\rightarrow 1$ the Abrikosov-Gorkov and for $\Gamma \rightarrow 0$
the BCS result is recovered. The  normalized density of states is
$\mathrm{Re}[G(E)]$. The rate of recombination with phonon emission
is \cite{kaplan},
\begin{eqnarray}
\label{equation:kaplan}
\frac{1}{\tau_{r}(\epsilon)}=&&\frac{1}{\tau_0 (kT_{c0})^3 [1-f(\epsilon)]} \int_{0}^{\infty} (E+\epsilon)^2 \Big(\mathrm{Re}[G(E)] \nonumber\\
&&+\frac{\Delta}{\epsilon} \mathrm{Im}[F(E)] \Big) [n(E+\epsilon)+1]f(E)dE
\end{eqnarray}
with $\tau_0$ denoting the material-dependent electron-phonon time,
assuming for the electron-phonon spectral function: $\alpha^2 F(E)
\propto E^2$, and $n(E)$ the phonon distribution function. On the
other hand, in the presence of pair-weakening $T_c$ and $\Delta$ are
reduced simultaneously, and the exponential dependence of the
recombination time on $T/T_c$ is retained. In Fig.
\ref{figure:MHZB}, the density of states (upper figure) and the
recombination time for quasiparticles at the minimum excitation
energy $\epsilon_{min}$ (lower figure) are shown for different
cases. Clearly, a density of states modified by magnetic impurities
results in a recombination time which remains temperature dependent,
independent of the model used.

We conclude that the recombination processes are unrelated to the
bulk magnetic moment of the implanted atoms, in agreement with the
observation that an enhancement can also be established by
implanting nonmagnetic atoms (Figs. \ref{figure:fig1},
\ref{figure:Tadata} and \ref{figure:Aldata}). Instead we attribute
the enhancement to an increase of the disorder caused by the
implantation. Impurities  might alter the electron-phonon
interaction \cite{rammer}, $\tau_0$ in Eq. \ref{equation:kaplan},
but no saturation would result \cite{barends}.

An interesting role of disorder, in particular at the surface, has
recently become apparent through phenomena controlled by unpaired
magnetic surface spins. An enhancement of the critical current of
nanowires has been observed \cite{rogachev}, in agreement with
theoretical predictions in which surface spins are aligned by the
magnetic field \cite{kharitonov}. In addition, recent tunneling
measurements on niobium surfaces show subgap states, Fig.
\ref{figure:MHZB}, signalling spins at the surface, possibly due to
the native oxide \cite{proslier}. Magnetic moments at surface
defects have also been proposed by Koch \textit{et al.} \cite{koch}
to explain the ubiquitous presence of flux noise in SQUIDs.
Sendelbach \textit{et al.} \cite{sendelbach} have observed in both
Al and Nb SQUIDs a strong dependence of the flux on temperature,
which they interpret as due to paramagnetic ordering of surface
spins by local fields in the vortex cores. In our recent experiments
on the frequency noise of superconducting resonators we also find a
strong dependence on the surface properties \cite{barendsAPL}. In
view of the other experiments, we conjecture that in our samples
unpaired surface spins are present, whose density is enhanced by the
ion bombardment. In order to properly address the relation to the
recombination rate, Eq. \ref{equation:kaplan} needs to be reanalyzed
taking into account spin flip \cite{grimaldi}, possible spin glass
formation \cite{sendelbach} and particle-hole asymmetry
\cite{yazdani}, giving rise to quasiparticles in the ground state
\cite{flatte}.

In conclusion, we have measured the relaxation time in Ta and Al
superconducting films implanted with both magnetic and nonmagnetic
impurities, using the complex conductivity. We find a clear trend of
decreasing relaxation time with increasing implanted impurity
concentration, independent of their magnetic moment. Our
observations show that low temperature quasiparticle recombination
is enhanced by disorder, most likely involving the surface.

\begin{acknowledgments}
The authors thank Y. J. Y. Lankwarden for fabrication of the
devices, K. van der Tak for the ion implantation, Ya. M. Blanter, T.
T. Heikkil\"{a} and Yu. V. Nazarov for stimulating discussions and
H. F. C. Hoevers for support. The work was supported by RadioNet
(EU) and the Netherlands Organisation for Scientific Research (NWO).
\end{acknowledgments}

\end{document}